\documentclass[aps,prc,twocolumn,showpacs,floatfix,nofootinbib,preprintnumbers,superscriptaddress,amsmath,amssymb]{revtex4-1}

\usepackage{epsfig}
\usepackage{hyperref}
\usepackage{ulem}
\usepackage{xcolor}

\begin{document}

\title{Tensor force role in $\beta$ decays analyzed within the Gogny-interaction shell model}

\author{B. Dai}
\affiliation{State Key Laboratory of
Nuclear Physics and Technology, School of Physics, Peking University, Beijing 100871,
China}
\author{B. S. Hu}
\affiliation{State Key Laboratory of
Nuclear Physics and Technology, School of Physics, Peking University, Beijing 100871,
China}
\author{Y. Z. Ma}
\affiliation{State Key Laboratory of
Nuclear Physics and Technology, School of Physics, Peking University, Beijing 100871,
China}
\author{J. G. Li}
\affiliation{State Key Laboratory of
Nuclear Physics and Technology, School of Physics, Peking University, Beijing 100871,
China}
\author{S. M. Wang}
\affiliation{State Key Laboratory of
Nuclear Physics and Technology, School of Physics, Peking University, Beijing 100871,
China}
\author{C. W. Johnson}
\affiliation{Department of Physics, San Diego State University, 5500 Campanile Drive, San Diego, California 92182-1233, USA} 
\author{F. R. Xu}\email{frxu@pku.edu.cn}
\affiliation{State Key Laboratory of
Nuclear Physics and Technology, School of Physics, Peking University, Beijing 100871,
China}


\begin{abstract}
\begin{description}
\item[Background]
The half-life of the famous $^{14}$C $\beta$ decay is anomalously long, with different mechanisms: the tensor force, cross-shell mixing, and three-body forces, proposed to explain the cancellations that lead to a small transition matrix element.
\item[Purpose]
We revisit and analyze  the role of the tensor force for the $\beta$ decay  of $^{14}$C as well as of neighboring isotopes.
\item[Methods]
We add a tensor force to the Gogny interaction, and derive an effective Hamiltonian for shell-model calculations. The calculations were carried out in a $p$-$sd$ model space to investigate cross-shell effects. Furthermore, we decompose the wave functions according to the total orbital angular momentum $L$ in order to analyze the effects of the tensor force and cross-shell mixing.
\item[Results] 
 The inclusion of the tensor force  significantly improves the shell-model calculations of the $\beta$-decay properties of carbon isotopes. In particular, the anomalously slow $\beta$ decay of $^{14}$C can be explained by the isospin $T=0$ part of the tensor force, which changes the components of $^{14}$N with the orbital angular momentum $L=0,1$, and results in a dramatic suppression of the Gamow-Teller transition strength. At the same time, the description of other nearby $\beta$ decays are improved. 
 
\item[Conclusions]
Decomposition of wave function into $L$ components illuminates how the tensor force modifies nuclear wave functions, in particular suppression of $\beta$-decay matrix elements. Cross-shell mixing also has a visible impact on the $\beta$-decay strength. Inclusion of the tensor force does not seem to  significantly change, however, binding energies of the nuclei within the phenomenological interaction.
\end{description}
\end{abstract}

\maketitle

\section{\label{sec:level1} Introduction}

Throughout the chart of the nuclides,  allowed $\beta$-decay lifetimes are typically short, e.g., the half-life of the $^{12}\text{N}$ $\rightarrow$ $^{12}$C $\beta$ decay is $\approx$ 11 ms.
The  $\beta$ decay  from the initial state $(J^{\pi}=0^+,T=1)$ in $^{14}$C to the final state $(J^{\pi}=1^+,T=0)$ in $^{14}$N, although allowed,  
 has a half-life of about 5730 y \cite{Ajzenberg1952,AJZENBERGSELOVE19911}.
This anomalously long half-life is not only useful in many fields \cite{Arnold678},
but also poses a challenge and an opportunity to test nuclear models and to understand fundamental interactions \cite{Ajzenberg1952,AJZENBERGSELOVE19911,Chou1993}. 

Many efforts \cite{Jancovici1954,Suzuki2003,Holt2008,Holt2009,Robson2011,Fayache1999,Maris2011,Ekstrom2014,Yuan2017} have been made to explain this puzzle. In the works \cite{Jancovici1954,Suzuki2003}, it was suggested that the tensor interaction plays an important role in the $\beta$ decay of $^{14}$C. Shell-model studies with realistic interactions in the $p$-shell supported this idea \cite{Holt2008,Holt2009,Robson2011,Fayache1999}.
More recent calculations within a larger $p$-$sd$ model space using the YSOX \cite{Yuan201212} and WBP \cite{Warburton1992} interactions found  off-diagonal cross-shell interactions between the $p$ and $sd$ shells lead to an anomalously small allowed matrix elements \cite{Yuan2017}. \textit{Ab initio}  calculations using the no-core shell model \cite{Maris2011} and coupled-cluster method \cite{Ekstrom2014} found that three-nucleon forces reduce the transition matrix element and dramatically increase the $^{14}$C lifetime. In both calculations, however, realistic three-body coupling constants yielded lifetimes an order of magnitude still too large.
This motivates us to revisit the tensor force.

In our previous work \cite{Jiang2018}, we have developed a shell-model calculation based on the Gogny interaction, and successfully applied it to the $p$-, $sd$- and $pf$-shell nuclei \cite{Jiang2018} and the $p$-shell hypernuclei \cite{Chen_2019}. However, the tensor force was excluded in the calculations.
In phenomenological potentials, e.g., Skyrme \cite{SKYRME1958615,Vautherin1972}, Gogny \cite{DECHARGE1975361,Decharge1980}, and relativistic mean field \cite{Roca-Maza2011}, the tensor force is usually not included explicitly. It has been recognized that the explicit inclusion of tensor force in mean-field calculations is nonetheless necessary for observables beyond ground-state (g.s.) energies \cite{COLO2007227,Lesinski200706}.
This has been shown in the mean-field calculations of $\beta$ decays \cite{Minato2013,BaiCL2014}. For these reasons, we have performed  calculations for carbon $\beta$ decays in the $p$-$sd$ space with the effective Hamiltonian derived from the Gogny interaction including a Gogny-type tensor force \cite{Otsuka2005}. 
We find that the inclusion of the tensor force not only reproduces the $^{14}$C half-life, it simultaneously improves the description of allowed transitions in nearby nuclides. 
Moreover, by decomposing the wave functions into orbital $L$ components \cite{Johnson2015}, we can clearly view the effect of the tensor force as well as cross-shell interactions on the wave functions.

\section{\label{sec:level2} The model}

\subsection{The Gogny interaction with tensor force embedded}

In shell-model calculations, the Hamiltonian is
\begin{eqnarray}
H &&=
\sum_{a} e_a \hat{n}_a + \sum_{a\leqslant b,c\leqslant d}\sum_{JT}V_{JT}(ab;cd)\widehat{T}_{JT}(ab;cd) ,
\label{eq1}
\end{eqnarray}
where $e_a$ and $\hat{n}_a$ are the energy and particle-number operator for the single-particle orbit $a$, respectively. $V_{JT}$ is the interaction two-body matrix elements (TBMEs), and $\widehat{T}$ is the two-body density operator for the nucleon pair in the orbits $(a,b)$ and $(c,d)$ with the coupled angular momentum $J$ and isospin $T$ \cite{Jiang2018}.

In the present work, we take the Gogny force \cite{DECHARGE1975361,Decharge1980} as the effective nucleon-nucleon interaction,
\begin{eqnarray}
V_{NN,12} =
	&&\sum_{i=1}^2 e^{-(\boldsymbol{r}_1-\boldsymbol{r}_2)^2/\mu_i^2} \notag\\
	&&\times (W_i+B_iP^{\sigma}-H_iP^{\tau}-M_iP^{\sigma}P^{\tau}) \notag\\
	&&+t_3\delta(\boldsymbol{r}_1-\boldsymbol{r}_2)(1+x_0P^{\sigma})\left[\rho(\frac{\boldsymbol{r}_1+\boldsymbol{r}_2}{2})\right]^\alpha \notag\\
	&&+iW_0\delta(\boldsymbol{r}_1-\boldsymbol{r}_2)(\boldsymbol{\sigma}_1+\boldsymbol{\sigma}_2)\cdot\boldsymbol{k}^{\prime}\times \boldsymbol{k},
	\label{eq2}
	\end{eqnarray}
where $P^\sigma = \frac{1}{2} (1+ \boldsymbol{\sigma}_1 \cdot \boldsymbol{\sigma}_2)$ and $P^\tau = \frac{1}{2}(1+\boldsymbol{\tau}_1 \cdot \boldsymbol{\tau}_2)$ are the spin- and isospin-exchange operators, and $\boldsymbol{\sigma}_i$ and $\boldsymbol{\tau}_i$ are the spin and isospin matrix vectors, respectively.
In the first term of Eq.\,(\ref{eq2}), the $\mu_i$ is the range of the Gaussian central force.
In addition, to consider the three-body effect on nuclear structure, a delta-type density-dependent effective three-body force has been included, where $\rho$ is the density of the nucleus at the center-of-mass (c.m.) position of the two interacting nucleons. We obtained the density $\rho$  through  self-consistent shell-model iterations \cite{Jiang2018}. The last term in the Gogny force is the spin-orbit coupling, where $\boldsymbol{k}=\frac{\overrightarrow{\nabla}_{1}-\overrightarrow{\nabla}_{2}}{2 i}$ and $\boldsymbol{k}^{\prime}=\frac{\overleftarrow{\nabla}_{1}-\overleftarrow{\nabla}_{2}}{2 i}$ are the relative wave vectors of the two nucleons.  We used  the D1S parameters \cite{BERGER1991365},  one of the most widely-used Gogny interactions. In the D1S, we have two ranges of $\mu_i$ = 0.7 and 1.2\,fm for the Gaussian central force, which represent the short-range repulsive and mid-range attractive behavior of the interaction.

We also add a tensor interaction similar to  Ref.\,\cite{ONISHI1978336}:
\begin{eqnarray}
V_{\rm{T}}(\boldsymbol{r}_1,\boldsymbol{r}_2)
	&& =\left[V_{\rm{T0}}(1-P^{\tau}_{12})-V_{\rm{T1}}(1+P^{\tau}_{12})\right] \notag\\
	&&~~\times S_{12}\text{exp}\!\!\left[-(\boldsymbol{r}_1-\boldsymbol{r}_2)^2/{\mu}^2_{\rm{T}}\right] \notag\\
	&& =\left[(\frac{1}{2}V_{\rm{T0}}-\frac{3}{2}V_{\rm{T1}})-(\frac{1}{2}V_{\rm{T0}}+\frac{1}{2}V_{\rm{T1}})\boldsymbol{\tau}_1\cdot\boldsymbol{\tau}_2\right] \notag\\
	&&~~\times S_{12}\text{exp}\!\!\left[-(\boldsymbol{r}_1-\boldsymbol{r}_2)^2/\mu^2_{\rm{T}}\right],
	\label{eq3}
\end{eqnarray}
where $S_{12}=3(\boldsymbol{\sigma}_1\cdot\boldsymbol{r})(\boldsymbol{\sigma}_2\cdot\boldsymbol{r})/r^2-\boldsymbol{\sigma}_1\cdot\boldsymbol{\sigma}_2$. 
In particular, we separate the tensor force into two channels: isospin $T=0$ singlet and $T=1$ triplet, with strengths $V_{\text{T0}}$ and $V_{\text{T1}}$, respectively. This allows us to study the effect of  the tensor force on  $\beta$ decay in detail.

\subsection{The Hamiltonian and shell-model calculations}\label{Sec:B}

With $^4$He as the inner core, we take the $p$-$sd$ shell as the model space for the calculations of carbon isotopes. TBMEs arise from the  D1S Gogny parametrization \cite{BERGER1991365},
while single-particle energies (SPEs) are chosen by fitting the experimental spectra of $^{15, 17}$O relative to the $^{16}$O ground state. Table\,\ref{SPEs} gives the SPEs used in the present $p$-$sd$ shell-model calculations. As shown in Table\,\ref{levels of O15 and O17}, the excited levels of $^{15, 17}$O can be well described using the SPEs given in Table\,\ref{SPEs} and TBMEs obtained with the D1S interaction. In the shell-model calculation, the harmonic-oscillator single-particle wave functions are used with a frequency parameter of $\hbar\omega=45A^{-1/3}-25A^{-2/3}$ MeV \cite{BLOMQVIST1968545}, and a full $p$-$sd$ model space is taken without further truncation. In the Hamiltonian, the Coulomb interaction is not included in order to keep the isospin symmetry, but its contribution to the binding energy (the ground-state energy) is considered as in Ref\,\cite{Yuan201212}. The Lawson method \cite{LAWSON1960963} is used to remove spurious c.m. motion.

\begin{table}
\caption{\label{SPEs} Single-particle energies (SPEs) used for the $p$-$sd$ shell-model space.}
\begin{ruledtabular}
\begin{tabular}{cccccc p{cm}}
  Orbit & $0p_{1/2}$ & $0p_{3/2}$ & $0d_{3/2}$ & $0d_{5/2}$ & $1s_{1/2}$\\
\colrule\\[-8pt]
SPEs (MeV) & 12.43 & 5.23 & 18.54 & 13.23 & 10.48\\
\end{tabular}
\end{ruledtabular}
\end{table}

\begin{table}
\caption{
\label{levels of O15 and O17}
Calculated spectra of $\rm{^{15}O}$ and $\rm{^{17}O}$ (relative to the $^{16}$O ground state) using the SPEs given in Table\,\ref{SPEs} and TBMEs obtained with the D1S interaction, compared with WBP \cite{Warburton1992} calculations and experimental data \cite{nndc}. 
}
\begin{ruledtabular}
\begin{tabular}{c|cccc p{cm}}
 Nuclei & $J^\pi$ & D1S (MeV) & WBP (MeV) &Exp (MeV)\\
\colrule\\[-8pt]
& $1/2^-$ & 15.88 & 17.61 & 15.66\\
$\rm{^{15}O}$ & $1/2^+$ & 20.23 & 24.83 & 20.85\\
&$5/2^+$ & 21.12 & 24.76 & 20.91\\
&$3/2^-$ & 22.55 & 22.97 & 21.84\\
\colrule\\[-8pt]
&  $5/2^+$ & -3.59 & -3.15 & -4.14\\
$\rm{^{17}O}$ &$1/2^+$ & -2.83 & -2.75 & -3.27\\
& $3/2^+$ & 0.44 & 2.42 & 0.94\\
\end{tabular}
\end{ruledtabular}
\end{table}

The tensor force given in Eq.\,(\ref{eq3}) takes a middle range of $\mu_{\rm{T}}=1.2$\,fm \cite{Otsuka2006,Anguiano2012}. The strengths of the $T=0$ and $T=1$ tensor forces are determined by fitting the experimental $B$(GT) values of $^{14}$C$(0^+_1)$ $\rightarrow$ $^{14}$N$(1^+_1)$, $^{12}$N$(1^+_1)$ $\rightarrow$ $^{12}$C$(0^+_1)$ and $^{12}$N$(1^+_1)$ $\rightarrow$ $^{12}$C$(2^+_1)$ $\beta$ decays (see details in Sec.\,\ref{sec:level3}), giving $V_{\text T0}=26.0$ MeV and $V_{\text T1}=34.8$ MeV. In this paper, we do not focus on the determinations of universal tensor strengths, but analyze the tensor force effect on the $\beta$ decays of nuclei around carbon.

\subsection{The $L$ decomposition for Gamow-Teller transition}

The $B(\text{GT})$ value of the Gamow-Teller (GT) transition is given by
\begin{eqnarray}
B(\text{GT})=\frac{1}{2J_i+1}|\langle J_f||\boldsymbol{\sigma}\boldsymbol{\tau}_{\pm}||J_i\rangle |^2,
\label{eq4}
\end{eqnarray}
where $J_i(J_f)$ is the angular momentum of the initial (final) state, and $\boldsymbol{\sigma}$ $(\boldsymbol{\tau})$ denotes the spin (isospin) operator.

Because the GT transition changes spin but does not change the spatial part of the wave function (i.e., the orbital angular momentum), the previous studies  \cite{Jancovici1954,Holt2008,Holt2009} dissected the $L$-$S$ structure  of the initial and final states. Those studies, however, relied upon the small dimensions of $p$-shell wave functions. 
Instead we use an efficient decomposition method suitable for large basis spaces~\cite{Johnson2015} to illuminate the effect of the tensor force on the wave functions and thus the transition.

The initial ($i$) and final ($f$) wave functions can be decomposed to different $L$ components,
\begin{eqnarray}
&&\left.| \Psi_i \right\rangle= a^i_{0} \left.| S_{i} \right\rangle +a^i_{1} \left.| P_{i} \right\rangle +   a^i_{2} \left.| D_{i} \right\rangle +\ldots, \\
&&\left.| \Psi_f \right\rangle= a^f_{0} \left.|S_{f}\right\rangle+a^f_{1} \left.|P_{f}\right\rangle+a^f_{2} \left.|D_{f}\right\rangle+\ldots, 
\label{Eq:L decomposition}
\end{eqnarray}
where $S,P,D\ldots$ are the components with good orbital angular momenta $L=0,1,2\ldots,$ respectively, and $a^i_{L},a^f_{L}$ are the $L$-decomposition coefficients of the initial and final wave functions, respectively. As indicated in Eq.\,(\ref{eq4}), the GT transition selects $\Delta L=0$ between the initial and final states. Therefore, we decompose the GT transition into different $L$ channels,  as
\begin{eqnarray}
\langle J_f||\boldsymbol{\sigma}\boldsymbol{\tau}_{\pm}||J_i\rangle &&=\sum_{L=0,1,2,\ldots}a^i_{L} a^f_{L} \langle L_f||\boldsymbol{\sigma}\boldsymbol{\tau}_{\pm}||L_i\rangle_{L_f=L_i=L}. \notag \\
\label{Eq:divided 1}
\end{eqnarray}
We define $M(\text{GT})=\langle J_f||\boldsymbol{\sigma}\boldsymbol{\tau}_{\pm}||J_i\rangle$ and $M_L(\text{GT})=\langle L_f||\boldsymbol{\sigma}\boldsymbol{\tau}_{\pm}||L_i\rangle_{L_f=L_i=L}$, then
\begin{eqnarray}
M(\text{GT})&&=\sum_{L=0,1,2,\ldots}a^i_{L}a^f_{L}M_L(\text{GT}) \notag\\
&&=\sum_{L=0,1,2,\ldots}M^{\rm{eff}}_{L}(\text{GT}),
	\label{Eq:divided GT}
\end{eqnarray}
where $M_L$(GT) is the strength of the GT transition for the $L$ channel, and $M_L^{\text{eff}}=a^i_{L}a^f_{L}M_L(\text{GT})$ is the effective transition strength containing the coefficients $a^i_{L}$ and $a^f_{L}$.

Usually, a quenching factor is used in $B$(GT) calculations \cite{BROWN2001517,Yuan201212,PhysRevLett.116.112502,10.1093/ptep/ptab022} to give better agreements with data, where Hamiltonians were determined normally by fitting data without including $B$(GT) values. In the present calculations, the tensor force strengths were determined by fitting the $B$(GT) data of $^{14}$C and $^{12}$N as mentioned in Sec. \ref{Sec:B}, without the quenching factor used in the fitting. Therefore, we do not use the quenching factor in the present $B$(GT) calculations.

\section{\label{sec:level3} Results and discussions}

\begin{figure}
\includegraphics[width=1\columnwidth]{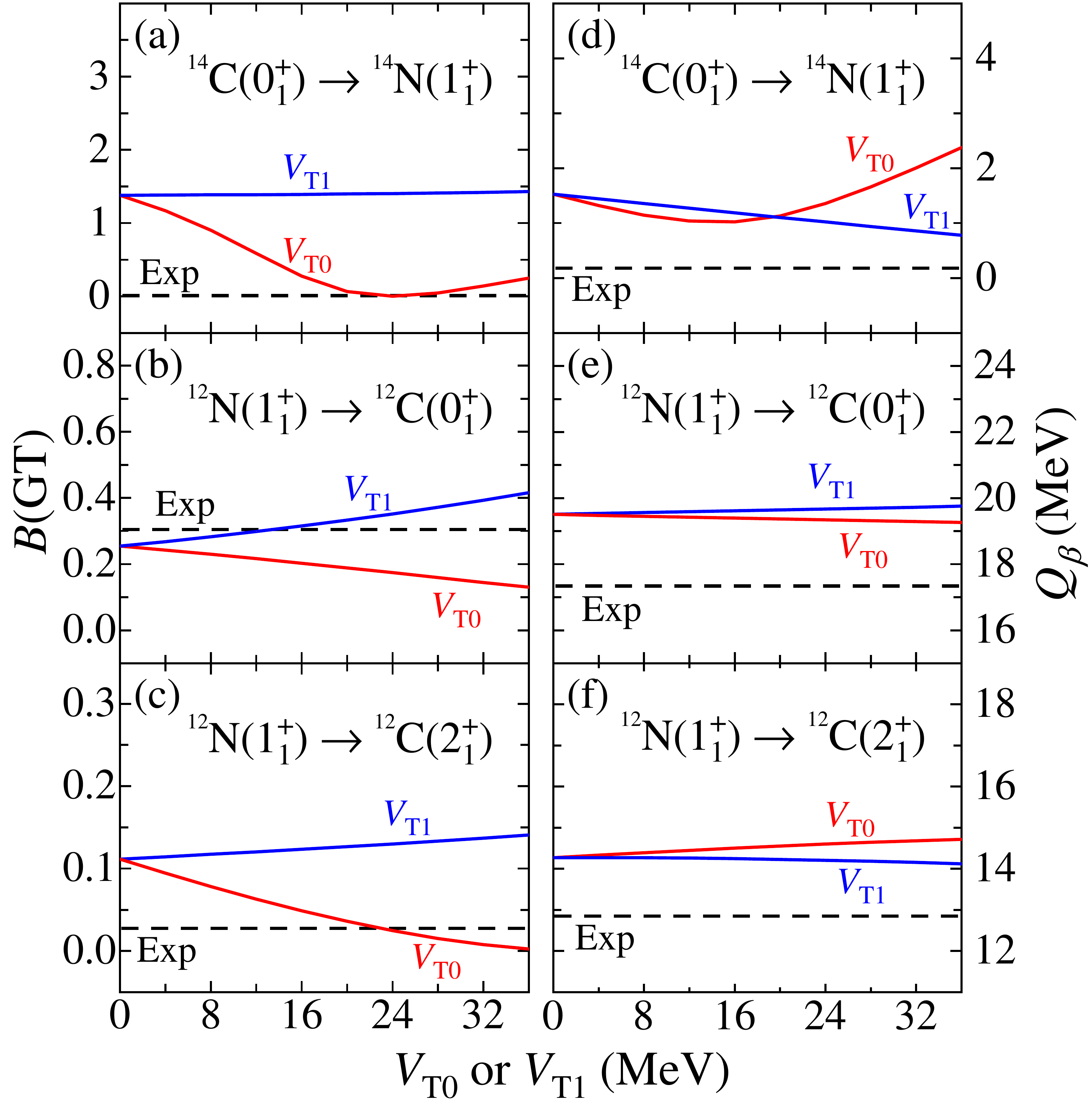}
\caption{\label{fig:Tensor value} Calculated $B$(GT) (left) and $Q$ (right) values of the $\beta$ decays of $\rm{^{14}C}$ $\rightarrow$ $\rm{^{14}N}$ and $\rm{^{12}N}$ $\rightarrow$ $\rm{^{12}C}$ as a function of the tensor strength $V_{\text{T0}}$ (in red) or $V_{\text{T1}}$ (in blue). The experimental data \cite{nndc} are indicated by black dashed lines.}
\end{figure}

\begin{figure}
\includegraphics[width=1\columnwidth]{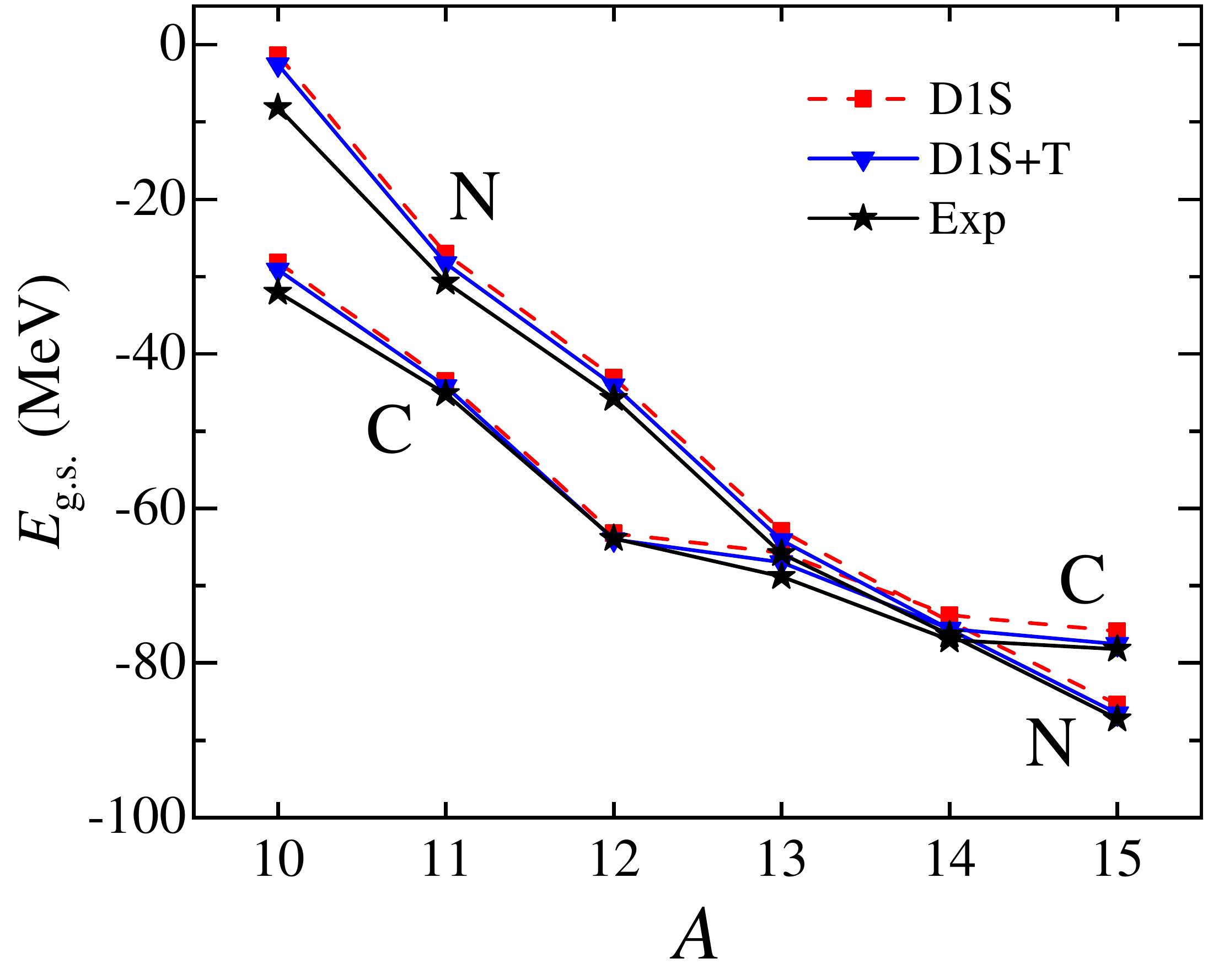}
\caption{Shell-model calculations of the ground-state energies (with respect to the $^4$He core) of carbon and nitrogen isotopes with (indicated by D1S+T) and without (indicated by D1S) the tensor force, along with experimental data \cite{ame}. 
\label{fig:Ground-state}}
\end{figure}

\begin{figure*}[hbt]
\includegraphics[width=1\textwidth]{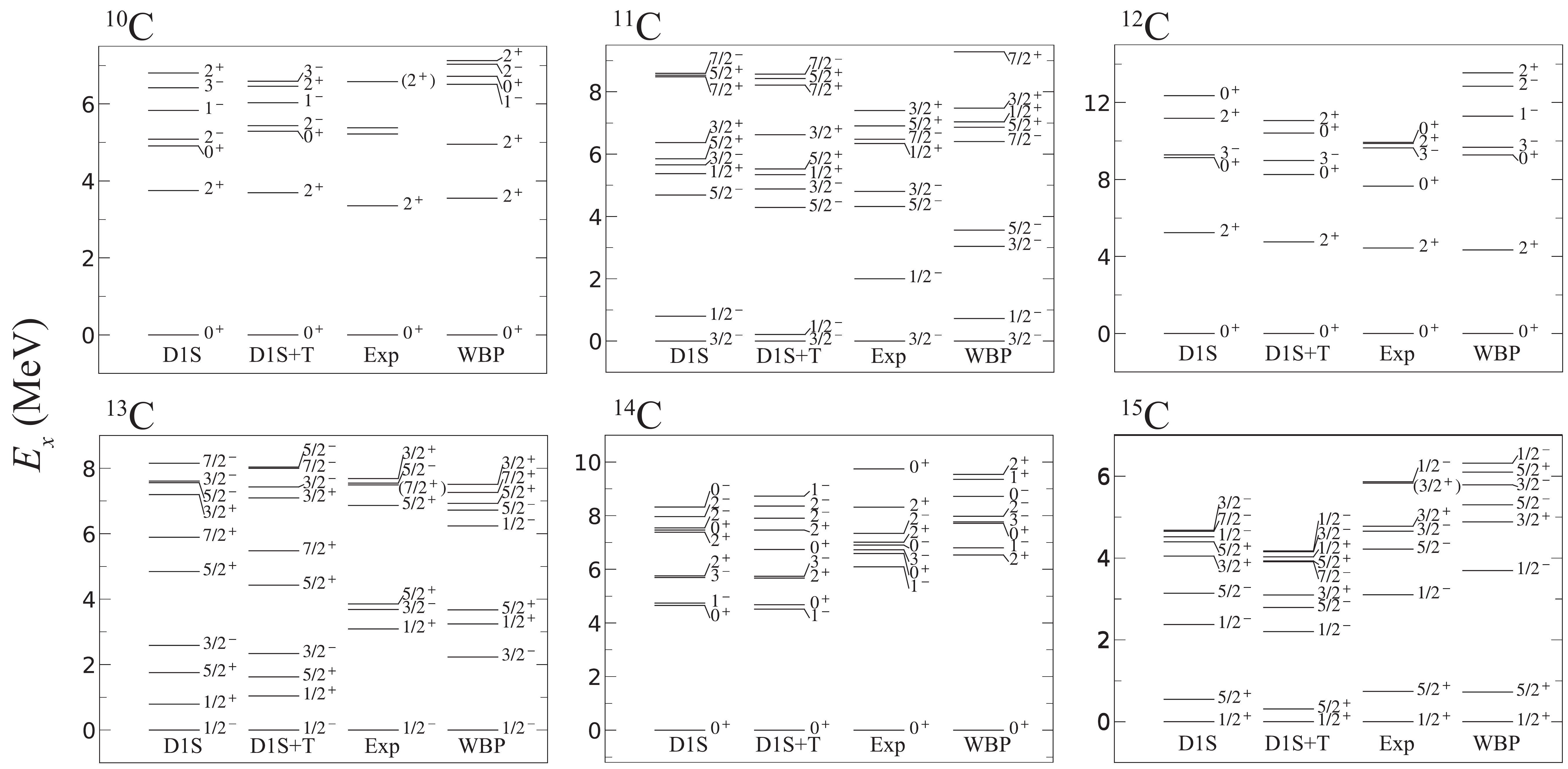}
\caption{\label{fig:levels} Shell-model calculations of spectra for carbon isotopes, with the effective interaction derived from the D1S Gogny interaction without and with the tensor force, indicated by D1S and D1S+T, respectively. The experimental data \cite{nndc} and calculations using the WBP interaction \cite{Warburton1992} are shown for comparisons.}
\end{figure*}

\begin{table*}
\caption{
\label{cases}
Calculated $B$(GT) values within the {\it p-sd} model space, using the D1S Gogny interaction embedded with different components of tensor force. T0 and T1 indicate the isospin $T=0$ and $T=1$ components of the tensor force, respectively. Also listed are  experimental data \cite{nndc} and the results with WBP interaction. 
}

\begin{ruledtabular}
\begin{tabular}{ccccccccc p{cm}}
  Transition & ($J_i^{\pi}$, $T_i$)&  ($J_f^{\pi}$, $T_f$)&  D1S & D1S+T0 & D1S+T1 & D1S+T1+T0 & WBP & Exp\\
\colrule\\[-8pt]
  $^{10}\rm{C}$ $\rightarrow$ $^{10}$B & ($0^+$, 1) & ($1^+$, 0) & 3.38 & 3.57 & 3.10 & 3.21 & 4.39 & 3.52\\
  $^{11}\rm{C}$ $\rightarrow$ $^{11}$B & ($\frac{3}{2}^-$, $\frac{1}{2}$) & ($\frac{3}{2}^-$, $\frac{1}{2}$) & 0.438 & 0.405 & 0.447 & 0.374 & 0.894 & 0.364\\
  $^{12}\rm{N}$ $\rightarrow$ $^{12}$C & ($1^+$, 1) & ($0^+$, 0) & 0.254 & 0.167 & 0.409 & 0.276 & 0.224 & 0.301\\
   & ($1^+$, 1) & ($2^+$, 0) & 0.112 & 0.0197 & 0.139 & 0.0288 & 0.00533 & 0.0276\\
  $^{13}\rm{N}$ $\rightarrow$ $^{13}$C & ($\frac{1}{2}^-$, $\frac{1}{2}$) & ($\frac{1}{2}^-$, $\frac{1}{2}$) & 0.376 & 0.304 & 0.356 & 0.254 & 0.298 & 0.211\\
  $^{14}\rm{C}$ $\rightarrow$ $^{14}$N & ($0^+$, 1) & ($1^+$, 0) & 1.38 & 0.0130 & 1.43 & $0.400\times 10^{-5}$ & 0.411 & $0.354\times 10^{-5}$\\
  $^{15}\rm{C}$ $\rightarrow$ $^{15}$N & ($\frac{1}{2}^+$, $\frac{3}{2}$) & ($\frac{1}{2}^+$, $\frac{1}{2}$) & 0.176 & 0.283 & 0.215 & 0.299 & 0.191 & 0.302\\
   & ($\frac{1}{2}^+$, $\frac{3}{2}$) & ($\frac{3}{2}^+$, $\frac{1}{2}$) & 0.327 & 0.0148 & 0.253 & 0.0530 & 0.0457 & $0.501\times 10^{-3}$\\
\end{tabular}
\end{ruledtabular}
\end{table*}

We computed the $\beta$-decay strength $B$(GT) and energy $Q_\beta$ of $^{14}\rm{C(0^+_{g.s.})}$ $\rightarrow$ $\rm{^{14}N(1^+_{g.s.})}$ and $\rm{^{12}N(1^+_{g.s.})}$ $\rightarrow$ $\rm{^{12}C(0^+_{g.s.})}$ while varying the strengths of the $T=0$ and $T=1$ components of the tensor force. As shown in Fig.\,\ref{fig:Tensor value}, $B$(GT) is sensitive to the strength of the tensor force, which agrees with previous work~\cite{Minato2013}. It is also interesting to notice that the effects of the $T=0$ and $T=1$ tensor forces are different in the allowed $^{14}$C $\rightarrow$ $^{14}$N and $^{12}$N $\rightarrow$ $^{12}$C $\beta$-decays. Figure \ref{fig:Tensor value}(a) shows that the $B$(GT) value of $^{14}\rm{C(0^+_{1})}$ $\rightarrow$ $\rm{^{14}N(1^+_{1})}$ is sensitive to $V_{\text T0}$ but insensitive to $V_{\text T1}$, 
while  Fig.\,\ref{fig:Tensor value}(b) shows that the $B(\text{GT})$ value of $\rm{^{12}N}({1^+_{1}})$ $\rightarrow$ $\rm{^{12}C}(0^+_{1})$ is sensitive to both of the  $T=0$ and $T=1$ tensor forces. From these two transitions, along with the $^{12}$N$(1^+_1)$ $\rightarrow$ $^{12}$C($2^+_1$) decay shown in Fig.\,\ref{fig:Tensor value}(c), we obtain $V_{\text T0}=26.0$ MeV and $V_{\text T1}=34.8$ MeV by fitting their experimental $B(\text{GT})$ values. The experimental $B(\text{GT})$ is obtained by
\begin{equation}
B(\text{F})+(\frac{g_{\text{A}}}{g_{\text{\text{V}}}})^2B(\text{GT})=\frac{K/g_{\text{V}}^2}{ft},
\label{Eq:GTtransition}
\end{equation}
where $K/g_{\text{V}}^2=6170$ \cite{Wilkinson1978} and $|g_{\text{A}}/g_{\text{V}}|=1.261$ \cite{Wilkinson1982} are taken, and $B({\text{F}})$ is the strength of the Fermi transition in the decay. In the present work, only $^{11}$C and $^{13}$N decays have the component of the Fermi transition with $B({\text{F}})=Z-N=1$ \cite{Suhonen2007}.

Our binding energies, and thus $Q_{\beta}$, are not sensitive to the tensor force, in agreement with prior density functional theory calculations \cite{COLO2007227}. Figure \ref{fig:Ground-state} shows the calculated ground-state energies with and without the tensor force (using our determined tensor force strengths) for the carbon and nitrogen chains, compared with data \cite{ame}. This insensitivity indicates that it is not necessary to refit the Gogny parameters with respect to binding energies when the tensor force is added; the main impact of the tensor force is in $\beta$-decay strengths.

Using these tensor force strengths, we have also calculated the $B$(GT) values for other carbon isotopes. Table\,\ref{cases} shows that the tensor force  remarkably improves the $\beta$-decay calculations compared with the D1S (without the tensor force included) and WBP predictions, especially for those with relatively small $B$(GT) values. This indicates that tensor force is essential to explaining systematics of  the $\beta$-decay strengths in this region, and that our interaction is consistent in this region of the nuclear chart. However, the tensor force effects are different for different decay systems. For example, the $T=0$ component enhances the $B$(GT) value in the $^{10}$C decay, while it suppresses the transition strengths of the $^{11,14}$C and $^{12,13}$N decays. Moreover, in some cases (e.g., in $^{10,14}$C decays), the $T=0$ and $T=1$ components have opposite contributions, similar to the decay of $^{12}$N.

Besides transition strengths and decay energies, we have also calculated the spectra of carbon isotopes (see Fig.\,\ref{fig:levels}). Results both with and without the tensor force agree reasonably with experimental spectra \cite{nndc} as well as those given with the WBP interaction \cite{Warburton1992}.  As with binding energies, the effect of the tensor force  on the spectrum is generally not significant. 
The tensor force is a rank-2 interaction for the orbital angular momentum, coupling configurations with $\Delta L \le 2$ \cite{Otsuka2005}; thus the magnitude of shifts in energies depends upon cross-shell configuration mixing in the $p$-$sd$ space.

\begin{figure}
\centering
\includegraphics[width=0.95\columnwidth]{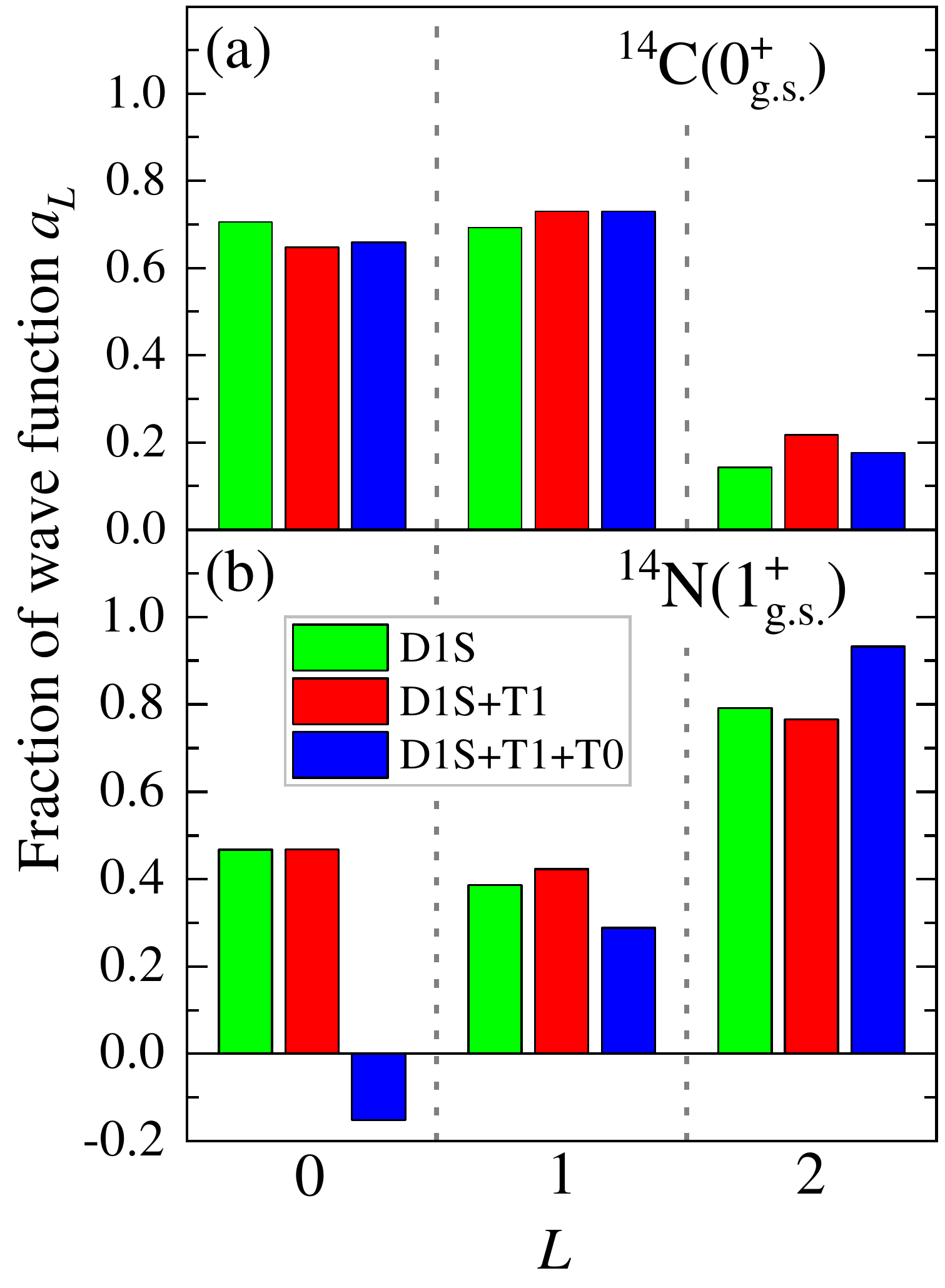}
\caption{\label{fig:L2_C14} $L$ decomposition for the g.s. wave functions of $^{14}$C (a) and $^{14}$N (b) in the $\beta$ decay of $^{14}$C$(0^+_{\text{g.s.}})$ $\rightarrow$ $^{14}$N$(1^+_{\text{g.s.}})$. The symbols of D1S, D1S+T1 and D1S+T1+T0 indicate the calculations with the D1S interaction only, the T1 tensor force added and both T1+T0 tensor forces included, respectively.}
\end{figure}

\begin{figure}
\includegraphics[width=0.95\columnwidth]{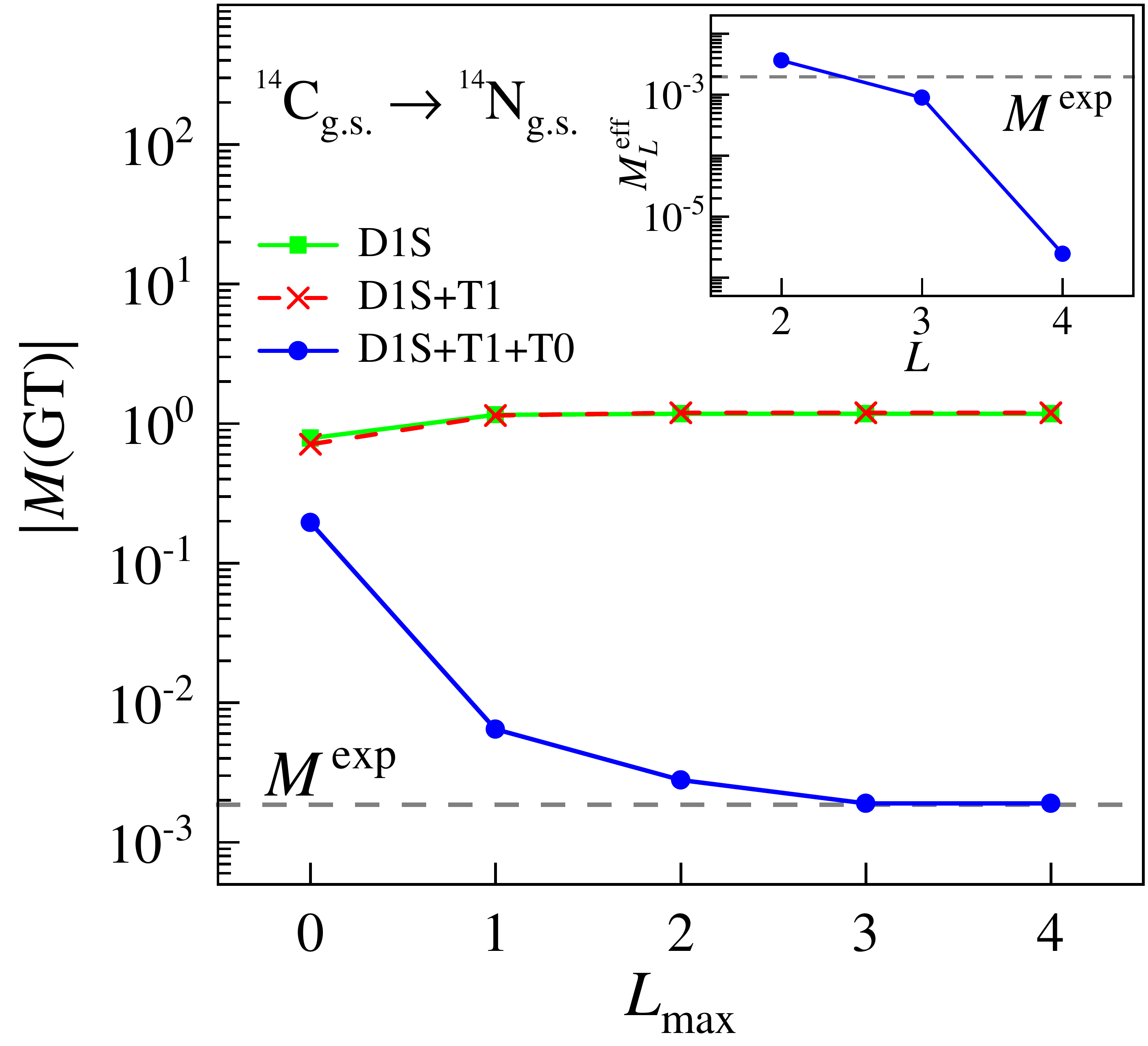}
\caption{\label{fig:L2_C14_2} The calculated GT transition strength $M(\text{GT})=\sum_{L=0}^{L_{\text{max}}} M_L^{\text{eff}}(\text{GT})$ for the $^{14}$C$(0^+_{\text{g.s.}})$ $\rightarrow$ $^{14}$N$(1^+_{\text{g.s.}})$ decay, with and without the tensor forces.  The experimental transition strength is extracted by $M^{\text{exp}}=\sqrt{(2J_i+1)B^{\text{exp}}({\text{GT}})}$ \cite{AJZENBERGSELOVE19911}.The insertion displays the calculated individual effective transition strength $M^{\text{eff}}_L$ at $L=2,3,4$, separately, showing the cross-shell effects.}
\end{figure}

To better understand the effects of the tensor force, we decompose the wave functions into components of total orbital angular momentum $L$.
Panels (a) and (b) of Fig.\,\ref{fig:L2_C14} show the $L$ decomposition for the $^{14}\rm{C}$ and $^{14}\rm{N}$ ground-state wave functions:  the $^{14}$C ground state is dominated by $L = 0,1$, while the $^{14}$N ground state is dominated by $L = 2$, consistent with other  $p$-shell calculations \cite{Jancovici1954}. 
In the decomposition, without considering the tensor force, we assign a positive sign to each $L$ component, i.e., assuming $a_L>0$. The tensor force may change the sign of the wave function. For example, as shown in Fig.\,\ref{fig:L2_C14}, the sign of the $L=0$ component in the $^{14}$N ground state becomes negative when the $T=0$ tensor force is considered. It is also shown that the $T=0$ tensor force reduces significantly the amplitudes of the $L=0,1$ components in $^{14}$N. 
As GT decay restricts $\Delta L=0$, this mismatch in $L$ between the initial and final states reduces the transition strength. The emergence of components with $L\geq 2$ in $^{14}$C$(0^+_{\text{g.s.}})$ and with $L\geq 3$ in $^{14}$N$(1^+_{\text{g.s.}})$ is due to cross-shell mixing between the $p$ and $sd$ shells.

In Fig.\,\ref{fig:L2_C14_2}, we plot the $^{14}$C$(0^+_{\text{g.s.}})$ $\rightarrow$ $^{14}$N($1^+_{\text{g.s.}}$) GT transition strength $M$(GT) defined by Eq.\,(\ref{Eq:divided GT}), as a function of the maximum angular momentum $L_{\text{max}}$ considered in the summation, i.e., 
\begin{equation}
M(\text{GT})=\sum_{L=0}^{L_{\text{max}}} M_L^{\text{eff}}(\text{GT}),
\end{equation}
where $M^{\text{eff}}_L({\text{GT}})$ is the transition strength for a fixed angular momentum $L$ as in Eq.\,(\ref{Eq:divided GT}). 
As shown in Fig.\,\ref{fig:L2_C14_2}, the transition strength drops dramatically from $L_{\text{max}}=0$ to $L_{\text{max}}=1$ with the $T=0$ tensor force included (see the blue curve). This is because the $T=0$ tensor force changes the sign of the $L=0$ wave function in $^{14}$N (see Fig.\,\ref{fig:L2_C14}), and then changes the relative sign between $M_{L=0}^{\text{eff}}$(GT) and $M_{L=1}^{\text{eff}}$(GT), which leads to a cancellation in the GT transition. 
Due this cancellation, the contributions from $L=2,3$ channels become competitive. The insert in Fig.\,\ref{fig:L2_C14_2} shows that the effective transition strengths $M_L^{\text{eff}}$ at $L = 2, 3$ are comparable to the experimental data, while higher-order effects from $L\geq 4$ channels are negligible.
 Note again that the $L\geq2$ channels can only come from the space beyond the $p$ shell, indicating the importance of cross-shell matrix elements. 

While we have focused on the coupling of different $L$ components via the tensor forces, another recent analysis focused on the role of isoscalar pairing \cite{utsuno2017noncoherent}, which can become incoherent depending on the relative sign of specific interaction matrix elements (in the case of the $^{14}$C GT transition, the $jj$-scheme matrix element $\langle p_{3/2} p_{1/2}|V|p_{1/2} p_{1/2}\rangle_{J=1, T=0}$), also leading to cancellations of GT matrix elements. In that analysis one source of the incoherence is the tensor force, consistent with both our work and that of Ref.\,\cite{Jancovici1954}.

\begin{figure}[hbt]
\includegraphics[width=0.95\columnwidth]{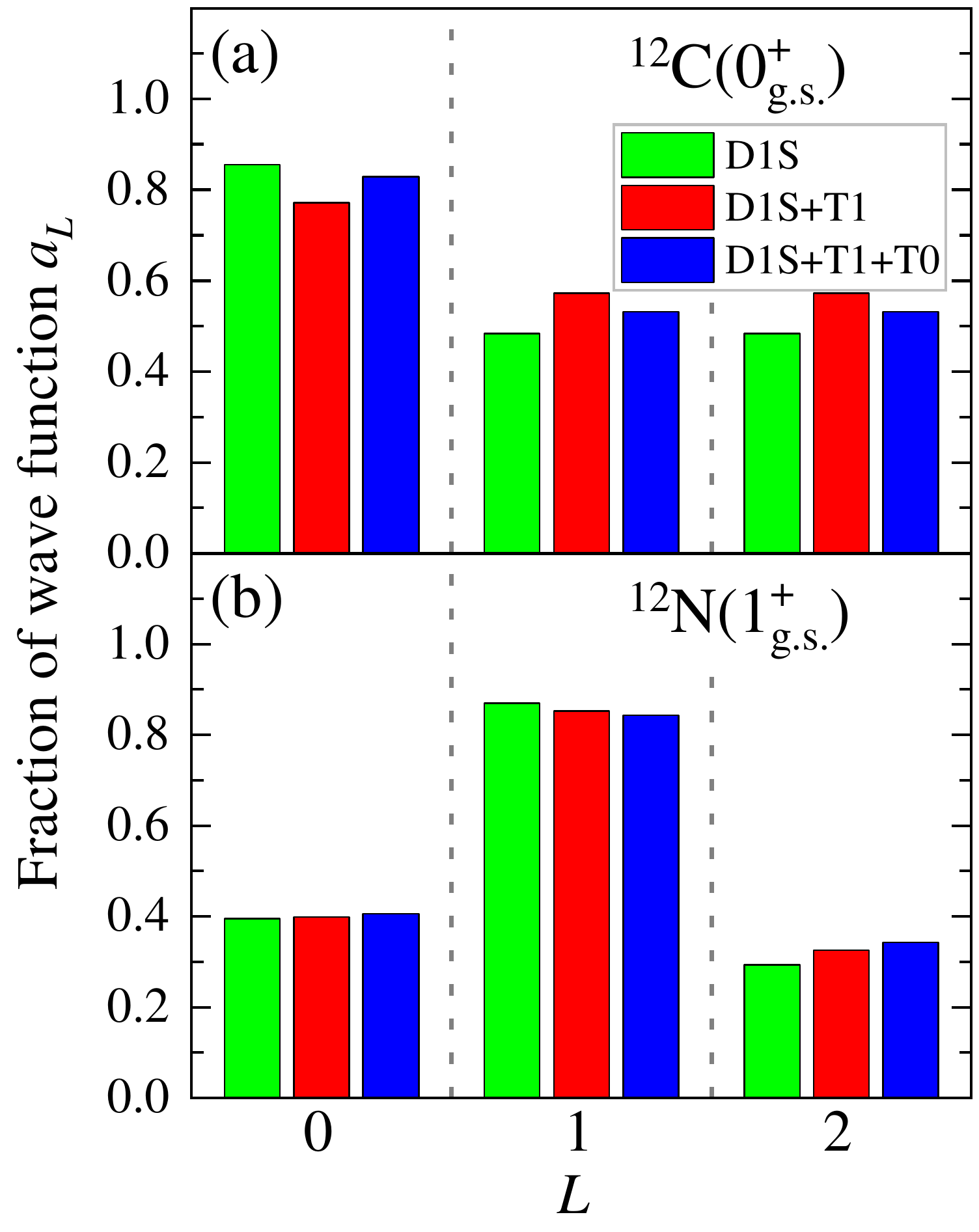}
\caption{\label{fig:L2_C12} Similar to Fig.\,\ref{fig:L2_C14}, but for the transition $^{12}$N$(1^+_{\text{g.s.}})$ $\rightarrow$ $^{12}$C$(0^+_{\text{g.s.}})$.}
\end{figure}

\begin{figure}[hbt]
\includegraphics[width=0.95\columnwidth]{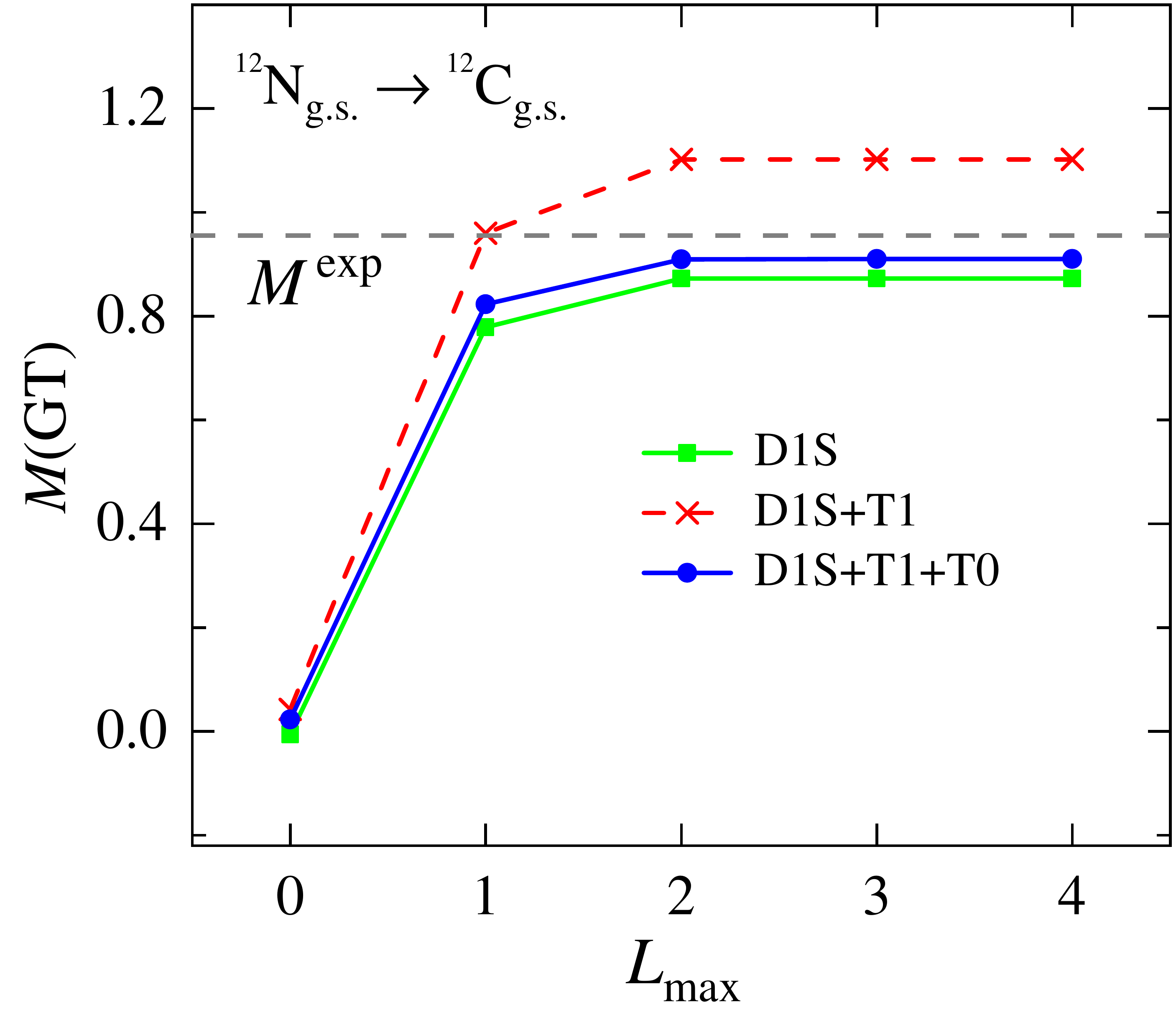}
\caption{\label{fig:L2_C12_2} Similar to Fig.\,\ref{fig:L2_C14_2}, but for the transition $^{12}$N$(1^+_{\text{g.s.}})$ $\rightarrow$ $^{12}$C$(0^+_{\text{g.s.}})$.}
\end{figure}

We have also investigated the GT transition $^{12}$N $\rightarrow$ $^{12}$C, which differs significantly from the decay of $^{14}$C.  The ground state of $\rm{^{12}C}$ has a dominant $L=0$ component, but  with a significant subdominant $L=1$ component, as seen in Fig.\,\ref{fig:L2_C12}(a), while the ground state of $^{12}$N is dominated by the $L=1$ component (about 70\% in the D1S calculation). Thus, unlike  $^{14}$N where the tensor force suppresses the $L=0,1$ components, this transition has significant overlap for $L=1$ which dominates the GT matrix element. This is another illustration of the usefulness of $L$ decomposition. Figure \ref{fig:L2_C12_2} shows that the tensor force has only a minor effect on the calculated GT transition strengths of the $^{12}$N $\beta$ decay.

\section{\label{sec:level6} Summary}

As an important ingredient of nuclear force, the tensor interaction has been shown to play a role in many observations of nuclear properties and processes. To study the tensor force and cross-shell effects on $\beta$ decays, we have calculated the Gamow-Teller transition strengths of  carbon isotopes using a shell-model framework with a tensor force embedded in the Gogny interaction.

We show that the tensor force plays a significant role in the $\beta$ decays of carbon isotopes, improving the description of not only the anomalous lifetime of $^{14}$C but also nearby nuclides. 
Furthermore, by decomposing the wave functions into components of total orbital angular momentum $L$  we can see more clearly the impact of the tensor force. The $T=0$ tensor force changes the sign of the $L=0$ component in the $^{14}$N ground state, which leads to a cancellation in the $^{14}$C $B$(GT) calculation between contributions from $L=0$ and $L=1$ channels. Consequently,
 cross-shell contributions from the $L=2,3$ channels play an important role in the $^{14}$C GT transition.  However, the $T=1$ component of the tensor force is useful to describe $\beta$ decays in nearby nuclides. 

While the roles of the tensor force and cross-shell matrix elements have been found in some previous calculations  \cite{Jancovici1954,Holt2008,Holt2009,Robson2011,Fayache1999,Yuan2017} , \textit{ab initio} calculations found an important role for three-nucleon forces \cite{Maris2011,Ekstrom2014}. 
In light of our results, it would be interesting to revisit such \textit{ab initio} frameworks and apply similar $L$ decompositions, as well as tracking 
Gamow-Teller matrix elements in nearby nuclides.  Such a comparison would be further the goal of understanding the origin of the anomalously long half-life of $^{14}$C.

\section*{\label{sec:level7} Acknowledgments}
This work has been supported by the National Key Research and Development Program of China under Grant No. 2018YFA0404401; the National Natural Science Foundation of China under Grants No. 11835001, 11921006, and 12035001; China Postdoctoral Science Foundation under Grant No. BX20200136; the State Key Laboratory of Nuclear Physics and Technology, Peking University under Grant No. NPT2020ZZ01; by the U.S. Department of Energy, Office of Science, Office of Nuclear Physics, under Award No. DE-FG02-03ER41272, and by the CUSTIPEN (China-U.S. Theory Institute for Physics with Exotic Nuclei) funded by the U.S. Department of Energy, Office of Science under Award No. DE-SC0009971. We acknowledge the High-Performance Computing Platform of Peking University for providing computational resources.

\normalem
\bibliography{references}

\end{document}